\documentclass[a4paper,12pt]{article}
\usepackage{amsmath}
\usepackage[latin1]{inputenc}
\usepackage{amssymb}
\usepackage{graphicx}
\usepackage{makeidx}
\usepackage{mathrsfs}
\usepackage{amsfonts}
\usepackage[mathscr]{eucal}
\usepackage{amsmath}
\font\fmtitle=cmbx12 scaled \magstep2
\font\text=cmr10 at 12 truept
 at 11 truept
 at 10 truept

\def\pni{\par \noindent}
\def\vsh{\vskip 0.25truecm\noindent}

\def\vsp{\vsh\pni}

\def\eg{{e.g.}\ }

\def\e{\hbox{e}}
\def\exp{\hbox{exp}}
\def\ds{\displaystyle}

\def\q{\quad}    

\def\l{\left} \def\r{\right}

\def\d{\partial}
   \def\dt{\partial t}
\def\dx{\partial x}     
\def\rec#1{\frac{1}{#1}}

\def\MM{{\rm I\hskip-2pt M}}
\def\RR{\vbox {\hbox to 8.9pt {I\hskip-2.1pt R\hfil}}\;}
\def\CC{{\rm C\hskip-4.8pt \vrule height 6pt width 12000sp\hskip 5pt}}

\def\exp{{\rm exp}\,} \def\e{{\rm e}}

\def\L{{\cal L}} 
\def\F{{\cal F}} 
\def\Fdiv{\,\stackrel{{\cal F}} {\leftrightarrow}\,}
  \def\Ldiv{\,\stackrel{{\cal L}} {\leftrightarrow}\,}
  \def\Mdiv{\,\stackrel{{\cal M}} {\leftrightarrow}\,}

\def\G{{\cal {G}}}

\begin{document}
\setcounter{page}{1}   \thispagestyle{empty}
\centerline{{\fmtitle The M-Wright function}}
\vskip 0.10truecm
\centerline{{\fmtitle in time-fractional diffusion processes:}}
\vskip 0.10 truecm
\centerline{{\fmtitle a tutorial survey}
\footnote{Paper published in
{\it International Journal of Differential Equations},
Vol. 2010, Article ID 104505, 29 pages. doi:10.1155/2010/104505 
in a special issue devoted to Fractional Differential Equations, see
{\tt http://www.hindawi.com/journals/ijde/2010/104505.abs.html}}}
\vskip 0.20truecm
\centerline{{\bf Francesco MAINARDI}$^{a}$, {\bf Antonio MURA}$^{b}$, 
and {\bf Gianni PAGNINI}$^c$ }
\noindent
\begin{center}
$^a$  Department of Physics, University of Bologna, and INFN,
 \\  Via Irnerio 46, I-40126 Bologna, Italy;
\\ E-mail: {\tt francesco.mainardi@unibo.it} \   
\vskip 0.05truecm
 $^b$  CRESME Ricerche S.p.A, 
\\ Viale Gorizia 25C, I-00199 Roma, Italy;
 \\ E-mail: {\tt anto.mura@gmail.com}
 \vskip 0.05truecm
$^c$  CRS4, Centro Ricerche Studi Superiori e Sviluppo in Sardegna,
\\ Polaris Bldg. 1,  I-09010 Pula (Cagliari), Italy;
\\ E-mail: {\tt pagnini@crs4.it}
\end{center}
\section*{Abstract}
\vskip -0.1truecm
In the present review we survey the properties of a transcendental function of the  Wright type,
 nowadays known as $M$-Wright function,
entering  as a probability density in  a  relevant class of self-similar 
stochastic processes that we generally refer to as  time-fractional diffusion processes.
 Indeed, the master equations governing these processes generalize the standard diffusion equation
 by means of time-integral operators interpreted as  derivatives of fractional order.   
When these generalized diffusion processes are properly characterized with stationary increments,
the $M$-Wright function is shown to play the same key role
  as the Gaussian density  in 
 the standard and fractional Brownian motions.
Furthermore, these processes  provide stochastic models suitable for describing 
phenomena of  anomalous diffusion of both slow and  fast type.
\section{Introduction}
\vskip -0.1truecm
 By time-fractional diffusion processes we mean certain diffusion-like phenomena
 governed by master equations containing fractional derivatives in time 
 whose fundamental solution
  can be interpreted as a probability density function ($pdf$) in space evolving in time.
  It is well known  that for the most elementary diffusion process,
 the Brownian motion, the master equation is the standard linear diffusion equation
 whose  fundamental solution is the Gaussian density  
 with a spatial variance growing linearly in time. In such case we speak about  normal diffusion,
 reserving the term anomalous diffusion when the variance grows differently.
 A number of stochastic models for explaining anomalous diffusion have been introduced
 in  literature, among them we like 
 to quote the fractional Brownian motion,  see \eg \cite{Mandelbrot-VanNess_68,Taqqu_REV02},
 the Continuous Time Random Walk, see 
 \eg \cite{GorMaiViv_CSF07,Meerschaert-et-al_PRE02,Metzler-Klafter_PhysRep00,Scalas_PRE04},
the L\'evy flights, see \eg  \cite{Dubkov-et-al_IJBC08},
 the Schneider grey Brownian motion, see \cite{Schneider_GN90a,Schneider_GN90b},
  and, more generally, random walk models based on evolution equations of single and distributed fractional order
 in time and/or space, 
 see \eg  \cite{Chechkin-Gorenflo-Sokolov_PRE02,Chechkin_FCAA03,Chechkin_PRE08},
 \cite{GorMai_NLD02,GorMai_CHEMPHYS02},
 \cite{Liu-et-al_ANZIAM05,Liu-et-al_AMC07},
 \cite{Zhang-et-al_JSP06,Zhuang-Liu_JAMC06}.
 \vsp
 In this survey paper we focus our attention  on  modifications of the standard diffusion	 equation,
 where the time can be stretched by a power law ($t \to t^\alpha$, $0< \alpha < 2$)
 and  the first-order time  derivative can  be replaced  by a  derivative of non-integer order 
 $\beta$ ($0< \beta\le 1$).
  In these cases of generalized diffusion processes  the corresponding fundamental solution still keeps  
  the meaning of a spatial $pdf$ evolving in time and is expressed   in terms 
  of a special function  of the Wright type that reduces to the Gaussian when $\beta=1$. 
  This transcendental function,  
  nowadays known as $M$-Wright function, will be shown  to play a fundamental role
   for a general class of self-similar stochastic processes with stationary increments, 
 which provide stochastic models for anomalous diffusion, as recently shown by   Mura et al.
\cite{Mura_PhD08,Mura-Mainardi_ITSF09,Mura-Pagnini_JPhysA08,Mura-Taqqu-Mainardi_PhysicaA08}.            
\vsp
In Section 2 we provide the reader with the essential notions and notations concerning 
the integral transforms and fractional calculus, which are necessary in the rest of the paper. 
In Section 3 we introduce in the complex plane $\CC$
the series and integral representations of the general Wright function
denoted by $W_{\lambda, \mu}(z)$ 
 and of the two related auxiliary functions $F_\nu(z)$, $M_\nu (z)$, which 
depend on a single parameter.  
In Section 4 we consider our  auxiliary functions in real domain pointing out 
their main properties involving their integrals and their asymptotic representations. 
Mostly, we restrict our attention to the second auxiliary function, that we call $M$-Wright function,
 when its variable is in $\RR^+$ or in all of $\RR$ but extended in symmetric way.
  We derive a fundamental formula  for the absolute moments of this function
in $\RR^+$, which allows us to obtain  its Laplace and Fourier transforms.
In Section 5 we consider some types of generalized diffusion equations  containing time partial derivatives
   of fractional order and we express their fundamental solutions in terms of the $M$-Wright functions
   evolving in time with a given self-similarity law.
   In Section 6 we stress how the  $M$-Wright function emerges as a natural generalization of the 
   Gaussian probability density for a class of  self-similar  stochastic processes with 
   stationary  increments, depending on two parameters ($\alpha, \beta$).
   These  processes are defined in a unique way by requiring the  determination of any multi-point
   probability distribution 
   and include the well-known standard and fractional Brownian motion.
   We refer to this class  as the generalized grey Brownian motion ($ggBm$), because it generalizes    
   the grey Brownian motion ($gBm$) introduced by Schneider \cite{Schneider_GN90a,Schneider_GN90b}. 
   Finally, a short concluding discussion    is drawn.  
   In Appendix A we derive the fundamental solution of the
   time-fractional diffusion equation. 
   In Appendix B we outline  the relevance of the $M$-Wright function 
   in time-fractional drift processes entering as subordinators in time-fractional diffusion.
     
\newpage
 \section{Notions and Notations}
  \paragraph{Integral transforms pairs.}
 \vsp
In our analysis we will make extensive use of integral transforms of Laplace, Fourier and Mellin
type so we 
first introduce our notation for the corresponding transform pairs. 
 We do not point out the conditions of validity and the main rules, since
they are given in any textbook on advanced mathematics. 
\vsp
Let
$$ \widetilde f(s) =
{\cal L} \left\{ f(r); r\to s\right\}
 = \int_0^{\infty} \e^{\ds \, -sr}\, f(r)\, dr\,,\eqno(2.1) $$
be the {\it Laplace transform} of   a  sufficiently well-behaved
function $f(r)$ with $r\in \RR^+$, $s\in \CC$, and let
$$ f(r) =
  {\cal L}^{-1} \left\{ \widetilde f(s);s \to r\right\}
 = \rec{2\pi i}\, \int_{Br}  
\e^{\ds \,+ sr}\, \widetilde f(s) \, ds\,,\eqno(2.2)  $$ be the inverse Laplace transform, where 
 $Br$ denotes the so-called Bromwich path, a straight line parallel to the imaginary axis
  in the complex $s$-plane.  
Denoting by $\Ldiv$ the justaposition of 
the original function $f(r)$ with its Laplace transform $\widetilde f(s)$, the Laplace transform pair reads
  $$ f(r) \,\Ldiv \, \widetilde f(s)\,.\eqno(2.3) $$
\vsp 
  Let
$$ \widehat f(\kappa) =
{\cal F} \left\{ f(x);x \rightarrow \kappa\right\}
 = \int_{-\infty}^{+\infty} \e^{\ds \,+ i\kappa x}\, f(x)\, dx\,,\eqno(2.4) $$
be the {\it Fourier transform} of   a  sufficiently well-behaved
function $f(x)$ with $x\in \RR$, $\kappa \in \RR$, and let
$$ f(x) =
  {\cal F}^{-1} \left\{ \widehat f(\kappa); \kappa \to x\right\}
 = \rec{2\pi }\, \int_{-\infty} ^{ + \infty}
\e^{\ds \, -i\kappa x}\, \widehat f(\kappa) \, d\kappa\,,\eqno(2.5)  $$
be the inverse Fourier transform. Denoting by $\Fdiv$ the justaposition of 
the original function $f(x)$ with its Fourier transform $\widehat f(\kappa)$, the Fourier transform pair reads
  $$ f(x) \,\Fdiv \, \widehat f(\kappa)\,. \eqno(2.6)$$
\vsp
Let
$$ f^*(s) =
{\cal M} \left\{ f(r);r \to s\right\}
 = \int_0^{\infty} r^{\ds \, s-1}\, f(r)\, dr\,,\eqno(2.7) $$
be the {\it Mellin transform} of   a  sufficiently well-behaved
function $f(r)$ with $r\in \RR^+$, $s\in \CC$, and let
$$ f(r) =
  {\cal M}^{-1} \left\{ f^*(s);s \to r  \right\}
 = \rec{2\pi i}\, \int_{Br}
r^{\ds \, -s}\,  f^*(s) \, ds\,, \eqno(2.8) $$
be the inverse Mellin transform. Denoting by $\Mdiv$ the justaposition of 
the original function $f(r)$ with its Mellin transform $f^*(s)$, the Mellin transform pair reads
  $$ f(r) \,\Mdiv \,  f^*(s)\,.\eqno(2.9) $$   
\paragraph{Essentials of fractional calculus with support in $\RR^+\,$.}
\vsp
Fractional calculus is the branch of mathematical analysis that deals 
with pseudo-differential operators that  extend the standard notions of 
integrals and derivatives to any positive non-integer order. 
The term fractional is kept only for historical reasons. Let us restrict our attention
to sufficiently well-behaved functions $f(t)$ with support in $\RR^+$.  
 Two main approaches  exist in the literature of fractional calculus to define 
the operator of derivative of non integer order for these functions,
referred to Riemann-Liouville and to Caputo. 
 Both approaches are related to the so-called Riemann-Liouville fractional integral 
 defined for any order $\mu>0$ as   
$$  J_t^\mu f(t) :=  \frac{1}{\Gamma(\mu)}\int_0^t (t-\tau)^{\mu-1} f(\tau) \, d\tau\,.
\eqno (2.10)$$
We note the convention $J_t^0 = I$ (Identity) and the semigroup property
$$ J_t^\mu \, J_t^\nu =  J_t^\nu \, J_t^\mu = J_t^{\mu + \nu}\,, \q
\mu \ge 0\,,\; \nu\ge 0\,.\eqno (2.11)$$
The fractional derivative of order $\mu >0$ in the {\it Riemann-Liouville} 
sense  is defined as the operator
$\,D_t^\mu$ which is the
left inverse of
the Riemann-Liouville integral of order $\mu $
(in analogy with the ordinary derivative), that is
$$ D_t^\mu \, J_t^\mu  = I\,, \q \mu >0\,. \eqno(2.12) $$
If $m$ denotes the positive integer
such that $m-1 <\mu  \le m\,,$  we recognize from Eqs. (2.11) and (2.12):
$\, D_t^\mu  \,f(t) :=  \, D_t^m\, J_t^{m-\mu}  \,f(t)\,, $
hence
$$
 D_t^\mu  \,f(t) = 
 \,
 \left\{
  \begin{array}{ll}
  {\ds \frac{d^m} {dt^m}}\left[
  {\ds \rec{\Gamma(m-\mu )}\int_0^t
    \frac{f(\tau)\,d\tau}  { (t-\tau )^{\mu  +1-m}} }\right] ,
 &  m-1 <\mu  < m, \\
   {\ds \frac{d^m} {dt^m} f(t)} \,,
&  \mu =m.
\end{array}
\right.
\eqno (2.13)$$
For completeness we define $ D_t^0 = I$.
\vsp
On the other hand, the fractional derivative of order $\mu >0$ in the
{\it Caputo} sense  is defined as the operator
$\,_*D_t^\mu$  such that
$    _*D_t^\mu \,f(t) :=  \, J_t^{m-\mu } \, D_t^m \,f(t)\,,$
hence
$$
    _*D_t^\mu \,f(t) =  
 \, \left\{
 \begin{array}{ll}
    {\ds \rec{\Gamma(m-\mu )}}\,{\ds\int_0^t
 {\ds \frac{f^{(m)}(\tau)\, d\tau}  {(t-\tau )^{\mu  +1-m}}}} \,,
&  m-1<\mu  <m,\\
   {\ds \frac{d^m}{dt^m} f(t)} \,, &  \mu =m.
   \end{array}
   \right.
\eqno(2.14)  $$
We note the different behavior of the two derivatives  in the limit $\mu \to (m-1)^+$. 
In fact, 
$$ \mu \to (m-1)^+ \; \left\{
\begin{array}{ll}
  & D_t^\mu f(t) \to  D_t^m \,J_t^1 \, f(t)= D_t^{(m-1)}\, f(t) \\
  & _*D_t^\mu f(t) \to  J_t^1 \, D_t^m \, f(t)= D_t^{(m-1)}\, f(t) - D_t^{(m-1)}f(0^+)\,,
   \end{array}
   \right.
   \eqno(2.15)$$ 
   where the limit for $t \to 0^+$ is taken after the operation of derivation.
 
 Furthermore, recalling the Riemann-Liouville fractional integral and derivative
of the power law for $t>0$,   
$$ 
\left\{
\begin{array}{ll}
& J_t^\mu \, t^{\gamma} = {\ds \frac{\Gamma(\gamma +1)}{\Gamma(\gamma +1+\mu)}\, t^{\gamma+\mu}\,,}\\
& D_t^{\mu}\, t^{\gamma}= {\ds \frac{\Gamma(\gamma +1)}{\Gamma(\gamma +1-\mu)}\, t^{\gamma-\mu}\,,}
\end{array} \right.     
 \q \mu >0\,,
  \; \gamma >-1\,, 
\eqno (2.16)$$
we find  the relationship between the two types of fractional derivative,
$$
   D^\mu \left[ f(t) -
 \sum_{k=0}^{m-1} \frac{t^k}{ k!} \, f^{(k)} (0^+)\right]
     =	\,_*D_t^\mu  \, f(t)  \,.\eqno(2.17)  $$
We note that the Caputo definition  for the
fractional derivative   incorporates the initial values
of the function and of its integer derivatives of lower order.
The subtraction of the Taylor polynomial of degree $m-1$ at $t=0^+$
from $f(t)$ is  a sort of
regularization	of the fractional derivative.
In particular, according to this definition,
the relevant property that the  derivative
of a constant is  zero is preserved for the fractional derivative.  
\vsp 
Let us finally point out the rules for the Laplace transform with respect to 
the fractional integral and the two fractional derivatives.
These rules are expected to properly generalize the well-known rules for standard integrals and derivatives.
\vsp
For the Riemann-Liouville fractional integral we have
$$ \L \left\{ J_t^\mu \,f(t) ;t \to s\right\} =
      \frac{ \widetilde f(s)}{s^\mu} \,,
  \q  \mu  \ge 0\,.\eqno(2.18) $$ 
For the Caputo fractional derivative we consequently get 
$$ \L \left\{ _*D_t^\mu \,f(t) ;t \to s\right\} =
      s^\mu \,  \widetilde f(s)
   -\sum_{k=0}^{m-1}    s^{\mu  -1-k}\, f^{(k)}(0^+) \,,
  \; m-1<\mu  \le m \,,\eqno(2.19) $$
where
$ f^{(k)}(0^+) := {\ds \lim_{t\to 0^+}}\, f^{(k)}(t)$.
The corresponding rule for the Riemann-Liouville
fractional derivative is more cumbersome and it reads
$$ \L \left\{ D_t^\mu  \, f(t);t \to s\right\} =
      s^\mu \,  \widetilde f(s)
   -\sum_{k=0}^{m-1}\,
\left[D_t^k\, J_t^{(m-\mu )}\right]\,f(0^+) \, s^{m -1-k}, \;  m-1<\mu  \le m ,
\eqno(2.20)$$
where the limit for $t \to 0^+$
is understood to be taken after the operations of fractional integration
and derivation.  As soon as all the limiting   values $f^{(k)}(0^+)$
are {\it finite}
and $m-1 <\mu< m$, 
 formula (2.20) for the Riemann-Liouville derivative  simplifies into
$$ \L \left\{ D_t^\mu  \, f(t);t \to s\right\} =
      s^\mu \,  \widetilde f(s) \,, \q m-1<\mu <m \,.\eqno(2.21)$$
In the special case   $f^{(k)}(0^+)=0$  for $k=0,1,  m-1$,
we recover the identity between the two fractional derivatives.
The Laplace transform rule (2.19)
was practically the key result of Caputo \cite{Caputo_GJRAS67,Caputo_BOOK69}
in defining his generalized derivative in the late sixties.
The two fractional derivatives have been  well discussed in the 1997 survey paper 
by Gorenflo and Mainardi \cite{GorMai_CISM97}, see also \cite{Mainardi-Gorenflo_FCAA07},
and in the 1999 book by Podlubny \cite{Podlubny_BOOK99}.
In these references the Authors have pointed out  
their preference  for the Caputo derivative in  physical applications
  where initial conditions are usually expressed in terms of
finite  derivatives of integer order.
 \vsp
For further reading 
on the theory and applications of fractional calculus
we recommend the recent treatise by 
Kilbas et al. \cite{Kilbas-et-al_BOOK06}.

\section{The functions of the Wright type}   
\paragraph*{The general Wright function.}
\vsp
The Wright function, that we denote by  $W_{\lambda,\mu,}(z)$, is so named in honour
of E. Maitland Wright, the eminent British mathematician, who introduced
and investigated this function
in a series of notes starting from 1933 in the framework of the asymptotic theory of partitions,
see  \cite{Wright_33,Wright_35a,Wright_35b}.
The function is defined by the series representation,
convergent in the whole $z$-complex plane,
  $$ W_{\lambda ,\mu }(z ) :=
   \sum_{n=0}^{\infty}\frac{z^n}{n!\, \Gamma(\lambda  n + \mu )}\,,
 \q \lambda  >-1\,, \; \mu \in \CC\,. \eqno(3.1)$$   
Originally, Wright assumed  $\lambda \ge 0$, and,
only  in 1940 \cite{Wright_40}, he considered 
$-1<\lambda <0$.
We note that in  Chapter 18 of Vol. 3 of the  handbook of the Bateman Project
\cite{Erdelyi_HTF},
devoted to Miscellaneous Functions, 
presumably for a misprint,  the parameter $\lambda $ of the Wright function is restricted to be non negative.
When necessary, we propose to distinguish the Wright functions in two kinds according to 
$\lambda \ge 0$ ({\it first kind})
and $-1<\lambda<0$ ({\it second kind}).
\vsp
For more details on  Wright functions the reader  can  consult \eg
 \cite{GoLuMa_99,GoLuMa_00,Kilbas-Saigo-Trujillo_02,Kiryakova_94,
 Mainardi_CISM97,Stankovic_70,Wong_99a,Wong_99b}
and references therein.
\vsp
The   {\it integral representation} of the Wright function reads
    $$ W_{\lambda ,\mu }(z )
 = \rec{2\pi i}\,\int_{Ha}   \!\!
 \e^{\, \ds \sigma +z\sigma ^{-\lambda }} \,
   \frac{d\sigma }{ \sigma^{\mu}} \,, 
    \q \lambda  >-1\,, \; \mu \in \CC\,,
   \eqno(3.2)$$   
where $Ha$ denotes the Hankel path.
We remind that the Hankel path is
a loop that starts from $-\infty$ along the lower side of the negative
real axis, encircles the circular area around the origin with radius $\epsilon \to 0$ 
in the positive sense, and ends at $-\infty$ along the upper side of the negative real axis.
The equivalence of the series and integral representations
is easily proved  using  Hankel formula for the Gamma
function
$$\rec{\Gamma(\zeta )} = \int_{Ha} \e^{\, \ds u} \, u^{-\zeta }\, du\,,
  \q \zeta \in \CC\,, $$
and performing a term-by-term integration.
In fact,
 $$
W_{\lambda ,\mu }(z) =
  \rec{2\pi i}\,\int_{Ha}   \!\!
 \e^{\, \ds \sigma +z\sigma ^{-\lambda }} \,
   \frac{d\sigma }{  \sigma^{\mu}}
 = \rec{2\pi i}\,\int_{Ha}   \!\! \e^{\,\ds \sigma}\,
 \l[\sum_{n=0}^{\infty}
 \frac{ z^n}{  n!} \,\sigma^{-\lambda  n}\r]\,
 \frac{d\sigma  }{  \sigma^{\mu }}  $$
$$= \sum_{n=0}^{\infty} \frac{ z^n  }{  n!}
 \l[ \rec{2\pi i} \int_{Ha} \!\!
 \e^{\, \ds \sigma} \, \sigma ^{-\lambda  n -\mu } \, d\sigma \r]
 = \sum_{n=0}^{\infty}
 \frac{ z^n}{  n!\, \Gamma[\lambda  n + \mu ]}\,.
 $$
It is possible to prove that the Wright function is entire of order
$1/(1+\lambda)\,, $ hence it is 
of exponential type only if $\lambda \ge 0$ (which corresponds 
to Wright functions  of the first kind).
The case $\lambda =0$ is trivial
since
$  W_{0, \mu }(z) = { \e^{\, z}/ \Gamma(\mu )}\,,$
provided that $\mu \ne 0, -1, -2, \dots$.
\paragraph{The auxiliary functions of the Wright type.}
\vsp
Mainardi, in his  first analysis of the time-fractional diffusion equation
  \cite{Mainardi_WASCOM93,Mainardi-Tomirotti_TMSF95},  
 aware of the Bateman handbook \cite{Erdelyi_HTF}, but not yet of the 1940 paper by Wright \cite{Wright_40},
introduced the two  (Wright-type) entire  {\it auxiliary functions},
$$ F_\nu (z) :=   W _{-\nu , 0}(-z)\,, \q 0<\nu<1\,, \eqno(3.3)$$
and
$$ M_\nu (z) :=  W _{-\nu , 1-\nu }(-z)\,,
\q 0<\nu<1 \,,\eqno(3.4)$$
inter-related through
$$ F_\nu (z) = \nu  \, z \, M_\nu (z ) \,.\eqno(3.5)$$
\vsp
As a matter of fact, 
 functions $F_\nu (z)$ and $M_\nu(z)$ are particular cases of the Wright function of the second kind
$W_{\lambda, \mu}(z)$
by setting $\lambda = -\nu$  and $\mu =0$ or $\mu=1$, respectively. 
\vsp
Hereafter, we provide the series and integral representations
of the two auxiliary functions derived from the general formulas (3.1) and (3.2), respectively.
\vsp
The {\it series representations} 
for the auxiliary functions  read
  $$ 
 F_\nu (z) :=
 {\ds \sum_{n=1}^{\infty}
\frac{(-z)^n}{  n!\, \Gamma(-\nu n)}}   
   = {\ds \rec{\pi}\,
\sum_{n=1}^{\infty}
 \frac{(-z)^{n-1}}{ n!}\,
 \Gamma(\nu n +1 )\, \sin(\pi \nu  n)\,,}
\eqno(3.6)   $$
 and
$$ 
M_\nu (z) :=
 {\ds \sum_{n=0}^{\infty}
 \frac{(-z)^n }{  n!\, \Gamma[-\nu n + (1-\nu )]} } 
  = {\ds \rec{\pi}\, \sum_{n=1}^{\infty}\,\frac{(-z)^{n-1} }{  (n-1)!}\,
  \Gamma(\nu n)  \,\sin (\pi\nu n)}  \,.
  \eqno(3.7)$$
The second series  representations in Eqs. (3.6)-(3.7) 
 have been obtained by using the reflection formula for the Gamma function
 $\, \Gamma(\zeta)\,\Gamma(1-\zeta)  =\pi /\sin\,\pi \zeta$.
 \vsp
As an exercise, the reader can directly prove that the radius of convergence of the power series 
in (3.6)-(3.7) is  infinite for $0<\nu<1$ without being aware of  Wright's results,
as it was shown independently by Mainardi \cite{Mainardi_WASCOM93}, see also \cite{Podlubny_BOOK99}. 
\vsp
Furthermore, we have  $F_\nu (0)= 0$
and $M_\nu (0) = 1/\Gamma(1-\nu)$.
We note that  relation (3.5) between the two auxiliary functions
can be easily deduced     from (3.6)-(3.7), by using the basic property of the Gamma function
 $ \Gamma(\zeta+1)= \zeta\,\Gamma(\zeta)$.
\vsp
The {\it integral representations} 
for the auxiliary functions  read 
$$ F_\nu (z) :=
 \rec{2\pi i}\int_{Ha}   \!\!  \e^{\ds \, \sigma -z\sigma ^\nu} 
   d\sigma \,,  \eqno(3.8)$$
$$   M_\nu (z ) :=
 \rec{2\pi i}\int_{Ha}   \!\!
 \e^{\ds \,\sigma -z\sigma ^\nu} 
   \frac{d\sigma}{  \sigma ^{1-\nu }}\,. \eqno(3.9)$$
We note that  relation  (3.5)
 can be obtained also from (3.8)-(3.9) with
an integration by parts. In fact, 
$$ \begin{array}{ll}
M_\nu(z) & ={\ds \int_{Ha}   \!\! \e^{\ds \,\sigma -z\sigma ^\nu} \,
   \frac{d\sigma}{  \sigma ^{1-\nu }}}
  = {\ds \int_{Ha}   \!\! \e^{\ds \,\sigma} \,
 \left(-\rec{\nu z}\, \frac{d}{  d\sigma } \e^{\ds\, -z\sigma ^\nu}\right) \,
   d\sigma} \\ \\
   & = {\ds \rec{\nu z}\,
   \int_{Ha}   \!\! \e^{\ds \, \sigma -z\sigma ^\nu} \, d\sigma}
   = {\ds \frac{F_\nu(z)}{\nu z}}
   \,.
   \end{array}
    $$
The equivalence of the series and integral representations
is easily proved by  using the Hankel formula for the Gamma
function and performing a term-by-term integration.	
	\vsp
\paragraph*{Special cases.}
\vsp
Explicit expressions of $F_\nu (z) $ and $M_\nu (z)$ in terms of known
functions are expected for some particular values of $\nu $.
Mainardi and Tomirotti \cite{Mainardi-Tomirotti_TMSF95}  
 have shown that for  $\nu =1/q\,,$
where $q \ge 2\,$ is a positive integer,
the auxiliary functions  can be expressed as a sum of
 simpler $(q-1)$ entire functions.
In  the particular cases $q=2$ and $q=3$ we find
$$ \!\!  M_{1/2}(z)\! =\! \rec{\sqrt{\pi}}\,
    \sum_{m=0}^\infty (-1)^m \, {\left(\rec{2}\right)}_m\,
  \frac{z^{2m}}{  (2m)!}
 \!=\! \rec{\sqrt{\pi}}\, \exp \left(-{\,z^2/ 4}\right),
\eqno(3.10)$$
and
$$ \begin{array}{ll}
M_{1/3}(z) &=
 {\ds \rec{\Gamma(2/3)}\,
   \sum_{m=0}^{\infty}{\left(\rec{3}\right)}_m\,\frac{z^{3m}}{  (3m)!}
- \, \rec{\Gamma(1/3)}\,
  \sum_{m=0}^{\infty} {\left( \frac{2}{  3}\right)}_m \,
 \frac{z^{3m+1}}{  (3m+1)!}}\\
&={\ds  3^{2/3} \, {\rm Ai} \left( {z/ 3^{1/3}}\right)} \,,
\end{array}
\eqno(3.11)
 $$
where $Ai$ denotes the {\it Airy function}.
\vsp
Furthermore, it can be  proved  
 that
$M_{1/q}(z)$  satisfies the   differential equa\-tion 
of order $q-1$
$$ \frac{d^{q-1}}{  dz^{q-1}} \, M_{1/q}(z) +
  \frac{(-1)^q}{  q}\, z\, M_{1/q}(z) =0\,, \eqno(3.12)$$
subjected to the $q-1$ initial conditions at $z=0$,
derived from (3.7),
$$ M_{1/q}^{(h)}(0) 
=\frac{(-1)^h}{\Gamma[(1-(h+1)/q]}
=   \frac{(-1)^h}{  {\pi}}\,\Gamma[(h + 1)/q]\,\sin [\pi\, (h+1)/q] 
  \,,   
   \eqno(3.13)$$
  with $h = 0,\,1,\, \ldots \, q-2$.
We note that, for $q\ge 4\,,$ Eq. (3.12) is  akin to the
{\it hyper-Airy}  differential equation  of order $ q-1\,, $
see \eg \cite{Bender-Orszag_BOOK87}.
Consequently, 
the auxiliary function
$M_\nu (z)$ could be considered  as a sort of  {\it generalized hyper-Airy
function}. However, in view of further applications in stochastic processes, we prefer to consider
it  as a natural (fractional) generalization of the Gaussian function,
similarly as the Mittag-Leffler function is known to be the natural (fractional) generalization of 
the exponential function. To stress the relevance of the auxiliary function $M_\nu(z)$, 
it was also suggested  the special name 
 {\it M-Wright function}, a terminology that has been followed in literature to some extent
\footnote{Some authors including Podlubny \cite{Podlubny_BOOK99}, 
Gorenflo et al. \cite{GoLuMa_99,GoLuMa_00}, Hanyga \cite{Hanyga_TF-PRSL02},
 Balescu \cite{Balescu_CSF07},
Chechkin et al. \cite{Chechkin_PRE08}, 
Germano et al. \cite{Germano-et-al_PRE09},
Kiryakova \cite{Kiryakova_09a,Kiryakova_09b} 
  refer to the $M$-Wright function as the {\it Mainardi function}.
It was  Professor Stankovi{\'c},
during the presentation of the paper by Mainardi and Tomirotti
\cite{Mainardi-Tomirotti_TMSF95}
at the Conference {\it Transform Methods and Special Functions, Sofia 1994},
who informed  Mainardi, being aware only of the Bateman Handbook \cite{Erdelyi_HTF}, 
that the extension for $-1<\lambda < 0$
had  been already made   just by Wright himself in 1940 \cite{Wright_40}, following
his previous papers published in the thirties. 
Mainardi, in the paper \cite{Mainardi-Gorenflo-Vivoli_FCAA05}
 devoted to the 80-th birthday of Prof. Stankovi{\'c},
used the occasion
to renew  his personal gratitude to Prof.
Stankovi{\'c}  for this earlier information that led
 him to study the original papers by Wright and work  (also in collaboration)
on the functions of the Wright type for further applications, see \eg
\cite{GoLuMa_99,GoLuMa_00} and \cite{Mainardi-Pagnini_AMC03}.
}.   
\vsp
Moreover, the analysis of the limiting cases $\nu=0$ and $\nu =1$ requires special attention.
For $\nu=0$ we easily recognize from the series representations (3.6)-(3.7):
$$ F_0(z) \equiv 0\,, \quad M_0(z) = \e^{\, \ds -z}\,.$$
The limiting case $\nu=1$ is singular for both the auxiliary functions as expected from the definition
of the general Wright function when $\lambda = -\nu =-1$. Later we will deal with this singular case
for the $M$-Wright function when the variable is real and positive. 
 \section{Properties and plots of the auxiliary Wright functions in real domain}
 Let us state some relevant properties of the auxiliary Wright functions,
 with special attention to the $M_\nu$ function in view of its role
  in time-fractional diffusion processes.
\paragraph{Exponential Laplace transforms.}
 \vsp
 We start with  the Laplace transform pairs  involving exponentials in the Laplace domain.
 These were derived by Mainardi in his earlier  analysis of the time fractional diffusion equation, 
 see \eg \cite{Mainardi_WASCOM93}, \cite{Mainardi_AML96},
 $$ 
\rec{r}\, F_\nu \left( 1/{r^\nu } \right) =
   \frac{\nu }{  r^{\nu +1}}\,  M_\nu \left( 1/{r^\nu } \right)\,\Ldiv\,
    \e^{\ds \,-s^\nu}\,, \q  0<\nu <1\,, \eqno(4.1)$$
$$	
\rec{\nu}\, F_\nu \left( 1/{r^\nu } \right) =
   \frac{1}{  r^{\nu}}\,  M_\nu \left( 1/{r^\nu } \right)\,\Ldiv\,
    \frac{\e^{\ds\, -s^\nu}}{s^{1-\nu}}\,, \q  0<\nu <1\,. \eqno(4.2)$$
 We note that the inversion of the Laplace transform  of the exponential $\exp (-s^\nu)$ is relevant 
 since it yields for any $\nu \in (0,1)$ the  unilateral {\it extremal stable densities}
 in probability theory, denoted   
 by $L_\nu^{-\nu}(r)$ in \cite{Mainardi_LUMAPA01}.
  As a consequence,  the non-negativity of both the
 auxiliary Wright functions when their argument is positive is  proved by 
 the Bernstein theorem\footnote{
 We refer to Feller's treatise  \cite{Feller_71}  for  Laplace transforms, 
 stable densities and Bernstein theorem.}.     
 The  Laplace transform pair in (4.1) has a long history starting from a formal result 
 by Humbert \cite{Humbert_45} in 1945,  of which Pollard \cite{Pollard_46} provided a rigorous proof
 one year later.
 Then, in 1959  Mikusi{\'n}ski \cite{Mikusinski_59} 
 derived a similar result on the basis  of his  theory of operational calculus. 
In 1975, albeit unaware of  the previous results,
Buchen and Mainardi \cite{Buchen-Mainardi_75} derived the result in a formal way.
We note that all the above authors   were not informed about the Wright functions.
 To our actual knowledge the former author who derived the  Laplace transforms  pairs (4.1)-(4.2) in terms of 
 Wright functions  of the second kind was   Stankovi\`c in 1970, see \cite{Stankovic_70}. 
\vsp
Hereafter we would like to provide  two independent
proofs of (4.1) carrying out
the inversion of $\exp (-s^\nu)\,, $ either by  the complex Bromwich integral
formula following \cite{Mainardi_WASCOM93}, or  by  the formal series method
following \cite{Buchen-Mainardi_75}. Similarly we can act for the Laplace transform
pair (4.2).
For the complex integral approach we deform the Bromwich path $Br$ 
into the Hankel path $Ha$, that is equivalent to the original path, and we set $\sigma = s r$.
Recalling  the integral representation (3.8) for the $F_\nu$ function and Eq. (3.5), we get  
$$
{\cal{L}}^{-1} \,\left[  \exp \left(\ds -s^\nu\right); s \to r \right] =
   \rec{ 2\pi i}\,\int_{Br}   \!\!
 \e^{\ds \, sr - s^\nu} \,   ds   =
   \rec{ 2\pi i\, r }\,\int_{Ha}   \!\!
 \e^{\ds \, \sigma -(\sigma/r) ^\nu} \,   d\sigma $$
$$ =  \rec{ r}\, F_\nu \left( 1/{r^\nu } \right)
  =  \frac{\nu }{  r^{\nu +1}}\,  M_\nu \left( 1/{r^\nu }\right)
 \,. $$
 \newpage
 \noindent
 Expanding in power series the Laplace transform and inverting  term by term, we formally get 
 $$
{\cal{L}}^{-1} \, \left[ \exp \left(\ds -s^\nu\right)\right] =
   \sum_{n=0}^\infty
 \frac{(-1)^n  }{  n!}\,
    {\cal{L}}^{-1} \, \l[ s^{\nu n}\r]
   =  \sum_{n=1}^\infty
  \frac{(-1)^n  }{  n!}\,
    \frac{ r^{-\nu n-1}}{  \Gamma(-\nu n)}$$
$$ = \rec{r}\, F_\nu \left( 1/{r^\nu } \right) =
   \frac{\nu }{  r^{\nu +1}}\,  M_\nu \left( 1/{r^\nu } \right) \,,
 $$
 where now we have used the series representation (3.6) for the function $F_{\nu}$
 along with the relationship formula  (3.5). 

\paragraph{Asymptotic representation for large argument.}
   \vsp
Let us  point out    the asymptotic behaviour 
of the function $M_\nu(r)$ when $r \to \infty$. 
Choosing as
a variable $r/\nu $ rather than $r$, the computation of the desired 
asymptotic representation  by the saddle-point approximation is straightforward.
Mainardi and Tomirotti \cite{Mainardi-Tomirotti_TMSF95}
have obtained
$$
\begin{array}{ll}
M_\nu (r/\nu ) 
&\sim
   a(\nu )\, r^{\ds{(\nu -1/2)/(1-\nu)}}
  \,
   \exp{\left[-{\ds b(\nu)\,r}^{\ds {1/(1-\nu)}}\right]},\\ \\
&  a(\nu) = {\ds \rec{\sqrt{2\pi\,(1-\nu)}}    >0} \,,  \q
  b(\nu) = {\ds \frac{1-\nu }{  \nu }    >0} \,. 
\end{array}
\eqno(4.3)$$
The above evaluation is consistent with the first term in
the asymptotic series expansion provided by Wright  with a cumbersome and formal procedure  
for his general function $W_{\lambda,\mu}$ when
$-1<\lambda<0$, see \cite{Wright_40}.
In 1999 Wong and Zhao have derived  asymptotic expansions of the Wright functions of 
the first and second kind
  in the whole complex plane following  a new method for smoothing Stokes' discontinuities, see
  \cite{Wong_99a,Wong_99b}, respectively.
 \vsp
 We note that, for $\nu=1/2$ Eq. (4.3) provides the exact result consistent with (3.10),
 $$ M_{1/2}(2r)= \rec{\sqrt{\pi}} \e^{\ds -r^2} \Leftrightarrow M_{1/2}(r)= \rec{\sqrt{\pi}} \e^{\ds -r^2/4}\,.
 \eqno(4.4)$$
 We  also note that in the  limit $\nu \to 1^-$ the function $M_\nu(r)$ tends to the
   Dirac generalized function $\delta(r-1)$, as can be recognized also from the Laplace transform pair
   (4.1).   
\paragraph{Absolute moments.} 
\vsp     
   From the above considerations we recognize that,
   for the $M$-Wright functions,  the following rule for absolute moments in $\RR^+$ holds
   $$   \int_0^\infty \!\! r^\delta M_\nu(r)\, dr =
   \frac{\Gamma(\delta+1)}{  \Gamma(\nu \delta+1)}\,,
    \q\delta >-1\,,\q 0\le \nu <1\,. \eqno(4.5)$$
  In order to derive  this fundamental result, 
   we proceed as follows on the basis  of the integral representation (3.9):
  $$
  \begin{array}{lll}
{\ds  \int_0^{\infty} \!\! r^\delta\, M_\nu (r)\, dr} &=
 {\ds \int_{0}^{\infty} \!\! r^\delta \left[\rec{ 2\pi i}\int_{Ha}\!\!\! \e^{\,\sigma -r\sigma^\nu}
 \frac{d\sigma}{  \sigma^{1-\nu}} \right]dr } \\ \\
 &= {\ds  \rec{ 2\pi i}\int_{Ha}\!\!\! \e^{\,\sigma}\,
  \left[\int_0^{\infty} \!\!\e^{\,- r\sigma^\nu }\, r^\delta\, dr \right]\,
  \frac{d\sigma  }{  \sigma^{1-\nu}} }\\ \\
   &={\ds \frac{\Gamma(\delta+1)}{ 2\pi i} \int_{Ha}\! \frac{\e^{\sigma}}{ \sigma^{\nu \delta+1}}  \, d\sigma   =
 \frac{\Gamma(\delta+1)}{  \Gamma(\nu \delta+1)} }\,.
 \end{array}  $$
 Above we have legitimized  the exchange between  integrals and  used  the identity
 $$\int_0^{\infty} \!\!\e^{\,- r\sigma^\nu }\, r^\delta\, dr = 
 \frac{\Gamma(\delta+1)}{(\sigma^\nu)^{\delta+1}}\,,$$
 along with the Hankel formula of the Gamma function.
Analogously, we can compute  all the moments  of  $F_\nu (r)$ in $\RR^+$.  
\paragraph{The Laplace transform of the $M$-Wright function.}
\vsp
Let  the Mittag-Leffler function be defined in the complex plane for any $\nu \ge 0$ by the  
 following  series and integral representation,  
 see \eg \cite{Erdelyi_HTF,Mainardi-Gorenflo_JCAM00},
 $$E_\nu (z) =\sum_{n=0}^\infty  \frac{z^n}{\Gamma (\nu n +1)} =
  \rec{2\pi i}\,
 \int_{Ha}
\frac{\zeta ^{\nu -1} \, \e^{\,\zeta} }{  \zeta ^\nu -z}\,
        d\zeta \,,  \q  \nu  >0\,,
 \; z\in \CC\,. \eqno(4.6) $$
 Such function is entire of order $1/\alpha$ for $\alpha>0$ and  reduces to the function
 $\exp (z)$ for $\nu >0$ and to $1/(1-z)$ for $\nu =0$.
We recall that the Mittag-Leffler function for $\nu >0$ 
plays fundamental roles in applications of fractional calculus
like  fractional relaxation and fractional oscillation, see 
\eg 
 \cite{Achar-et-al_PhysicA04},
\cite{GorMai_CISM97}, \cite{Mainardi-Gorenflo_FCAA07}, \cite{Mainardi_BOOK2010},
so that it could be referred  as  the {\it  Queen function of fractional calculus}\footnote{%
Recently,  numerical routines for functions of  Mittag-Leffler type have
been provided 
\eg by  Freed et al. \cite{Freed_NASA02},   Gorenflo et al. \cite{Gorenflo-Loutchko-Luchko_FCAA02}
(with {\it MATHEMATICA}),  Podlubny \cite{Podlubny_MATLAB06}
(with {\it MATLAB}), Seybold and Hilfer   \cite{Seybold-Hilfer_FCAA05}.}. 
\vsp
 We now point out that the $M$-Wright function is related to the Mittag-Leffler function
through the following Laplace transform pair,
$$ M_\nu (r)\, \Ldiv \, E_\nu (-s)\,, \q 0<\nu <1\,. \eqno(4.7)$$  
For the reader's  convenience we provide a simple proof of (4.7) by using two different approaches.
We assume that the exchanges between integrals and series are legitimate  in view of the   
 analyticity properties of the involved functions.
In the first approach we use the integral representations of the two functions obtaining
$$ 
\begin{array}{lll}
{\ds  \int_0^\infty \e^{\ds\, -sr} \, M_\nu(r)\, dr} &=  
{\ds \rec{ 2\pi i}\,\int_0^\infty \e^{\ds\, -s\, r} \,
  \left[ \int_{Ha}   \!\! \e^{\ds \sigma -r\sigma ^\nu} \,
   \frac{d\sigma }{  \sigma^{\, \ds 1-\nu}}\right]\, dr} \\ \\
& ={\ds  \rec{ 2\pi i}\,\int_{Ha}   \!\! \e^{\,\ds \sigma} \,
    \sigma^{\, \ds \nu-1}\,\left[
 \int_0^\infty \e^{\ds\, - r(s+\sigma ^\nu )} \, dr \right]\, d\sigma} \\ \\
& = {\ds \rec{2\pi i}\,\int_{Ha}   \!\!
 \frac{ \e^{\ds \,\sigma} \, \sigma^{\, \ds \nu-1} }{  \sigma ^{\, \ds \nu} +s}\, d\sigma =E_\nu(-s)}
 \,.
 \end{array} 
 \eqno(4.8)$$
 In the second approach we develop in series the exponential kernel of the Laplace transform and
  we use the expression (4.5) for the absolute moments of the $M$-Wright function 
 arriving to the following series representation 
 of the  Mittag-Leffler function,
  $$ 
\begin{array}{ll}
{\ds  \int_0^\infty \e^{\ds\, -sr} \, M_\nu(r)\, dr} &=  
{\ds \sum_{n=0}^\infty  \frac{(-s)^n}{n!} \, \int_0^\infty  r^n \, M_\nu(r) \,dr}\\
&= {\ds \sum_{n=0}^\infty  \frac{(-s)^n}{n!}  \, \frac{\Gamma(n+1)}{\Gamma(\nu n+1)}
  =  \sum_{n=0}^\infty  \frac{(-s)^n}{\Gamma(\nu n+1)} = E_\nu(-s)}\,.
 \end{array} 
 \eqno(4.9)$$
 We note that the transformation term by term of the series expansion of the 
 $M$-Wright function is not legitimate
 because the function is not of exponential order, see \cite{Doetsch_BOOK74}. 
 However, this procedure yields the formal asymptotic expansion of the Mittag-Leffler function 
 $E_\nu(-s)$ as $s\to \infty$
 in a sector around the positive real axis, see \eg \cite{Erdelyi_HTF,Mainardi-Gorenflo_JCAM00}, that is
 $$ \begin{array}{ll}
 {\ds \sum_{n=0}^\infty \frac{\int_0^\infty \e^{\ds\, -sr} (-r)^n\, dr }{n! \Gamma(-\nu n +(1-\nu))}}
 &=
{\ds  \sum_{n=0}^\infty \frac{(-1)^n}{\Gamma(-\nu n +1-\nu)} \rec{s^{n+1}}} \\
& = {\ds  \sum_{m=1}^\infty \frac{(-1)^{m-1}}{\Gamma(-\nu m +1)} \rec{s^m}}
\sim E_\nu(-s)\,,\; s\to \infty\,. 
\end{array}
$$
\paragraph{The Fourier transform of the symmetric $M$-Wright function.}
\vsp
The $M$-Wright function, extended on the negative real axis as an even function, 
is related to the Mittag-Leffler function
through the following Fourier transform pair
$$ M_\nu (|x|)\, \Fdiv \,2 E_{2\nu} (-\kappa^2)\,, \q 0<\nu <1\,. \eqno(4.10)$$ 
Below, we  prove the equivalent formula
$$ \int_0^\infty \cos (\kappa r)\, M_\nu(r) \, dr = E_{2\nu} (-\kappa^2)\,.\eqno(4.11)$$
For the prove it is sufficient to develop in series the cosine function and use  formula (4.5)
for  the absolute moments of the  $M$-Wright function,
$$ \begin{array}{ll}
{\ds \int_0^\infty \cos (\kappa r)\, M_\nu (r) \, dr} &=
{\ds \sum_{n=0}^\infty (-1)^n \frac{\kappa^{2n}} {(2n)!}\, \int_0^\infty \!\!r^{2n}\, M_\nu(r)\, dr }\\
&= 
{\ds \sum_{n=0}^\infty (-1)^n \frac{\kappa^{2n}} {\Gamma(2\nu n +1)} = E_{2\nu}(-\kappa^2)}\,.
 \end{array}
 \eqno(4.12)
 $$      
\paragraph{The Mellin transform of the $M$-Wright function.}
\vsp
It is straightforward to derive the Mellin transform of the $M$-Wright function using  result (4.5)
for the absolute moments of the $M$-Wright function. 
In fact, setting $\delta =s-1$ in (4.5),  by analytic continuation it follows 
$$ M_\nu (r)\, \Mdiv \, \frac{\Gamma(s)}{\Gamma(\nu(s-1)+1)}\,, \q 0<\nu <1\,. \eqno(4.13)$$
\paragraph{Plots  of the symmetric $M$-Wright function.}  
\vsp
It is instructive to show the plots of the (symmetric) $M$-Wright function on the real axis 
   for some rational values of the parameter $\nu$.  In order to have more  insight
   of the effect of the parameter itself on the behaviour close to and far from the origin,
     we  adopt both linear and logarithmic scale for the ordinates.
	 \vsp
	 In Figs. 1 and 2   
we compare the plots of the  $M_\nu (x)$-Wright functions
in $-5 \le x \le 5$ for some rational values of $\nu$ in the ranges  $\nu \in [0,1/2]$
and $\nu \in [1/2, 1]$, respectively.
In Fig. 1 
 we see the transition from $\exp (-|x|)$ for $\nu=0$
to $1/\sqrt{\pi}\, \exp (-x^2)$ for $\nu=1/2$, whereas
in Fig. 2 we 
see the transition from   $1/\sqrt{\pi}\, \exp (-x^2)$ 
to the delta functions $\delta(x\pm 1)$ for $\nu=1$.
Because of the two symmetrical humps  for $1/2<\nu\le 1$, the $M_\nu$ function appears bi-modal 
with the characteristic shape of the capital letter $M$.
\vsp	 
In plotting $M_\nu (x)$ at fixed $\nu $ for sufficiently large $x$
the asymptotic representation (4.3)-(4.4) is  useful
 since, as $x$ increases,
the numerical convergence of the series in (3.7)
decreases 
up to being completely inefficient:
henceforth, the matching between the series and the asymptotic
representation  is relevant and followed  by 
Mainardi and associates,  see \eg 
\cite{Mainardi_CHAOS96,Mainardi_CISM97, Mainardi_LUMAPA01,Mainardi-Pagnini_AMC03}.
However, as $\nu \to 1^-$,
the plotting remains a very difficult task because
of the high peak arising around $x= \pm 1$.
For this  we refer the reader to the 1997 paper by Mainardi and Tomirotti
\cite{Mainardi-Tomirotti_GEO97}, 
where a variant of the saddle point method has been successfully   used to properly depict the transition
to the delta functions $\delta(x \pm 1)$ as $\nu$ approaches  1.
For the numerical point of view we like to highlight the recent paper
by Luchko \cite{Luchko_FCAA08}, where 
algorithms are provided for computation of the Wright function on the real axis with
prescribed accuracy.
\vsp
\begin{figure}[ht!]
\begin{center}
 \includegraphics[width=.45\textwidth]{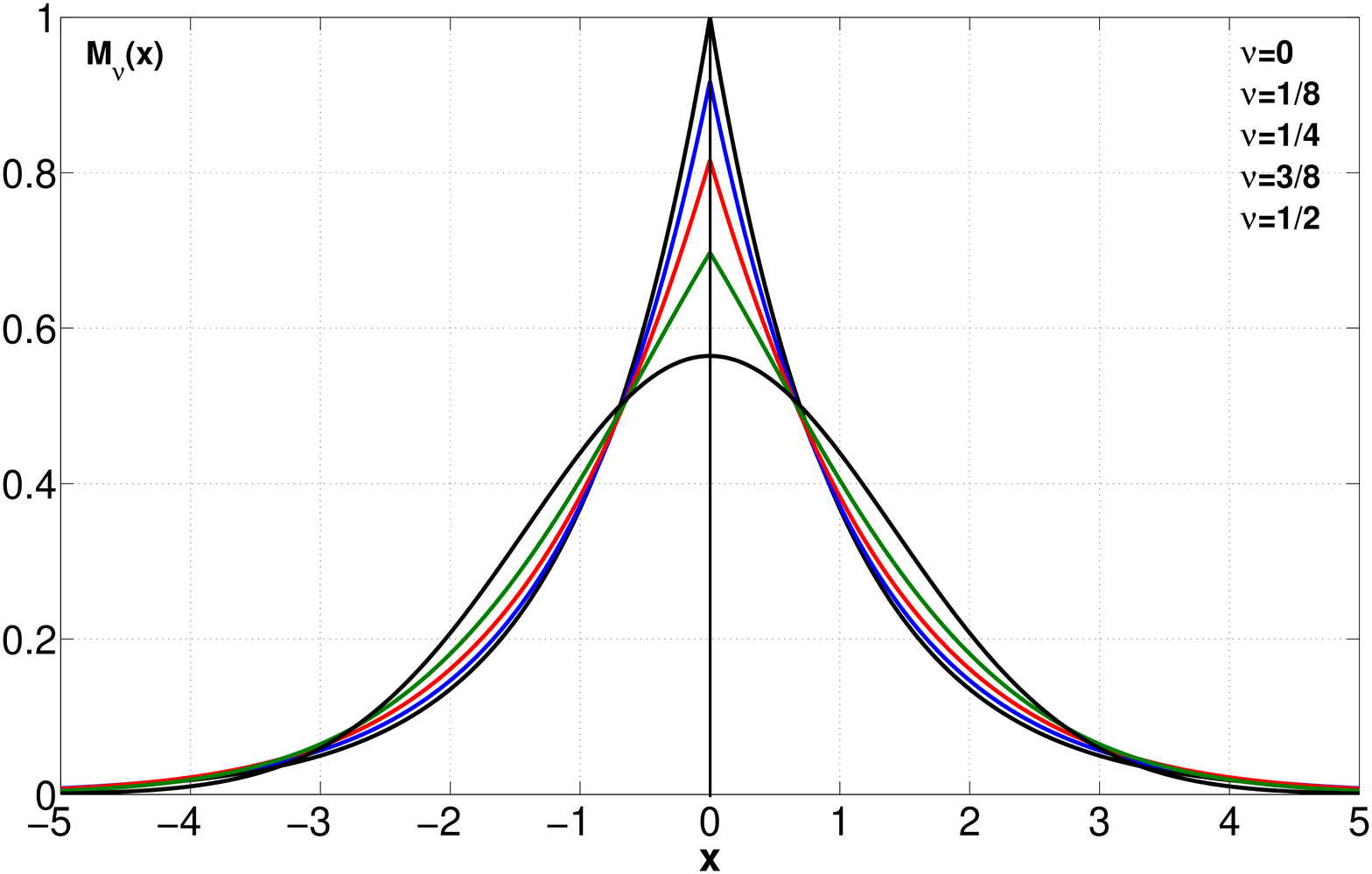}
 \includegraphics[width=.45\textwidth]{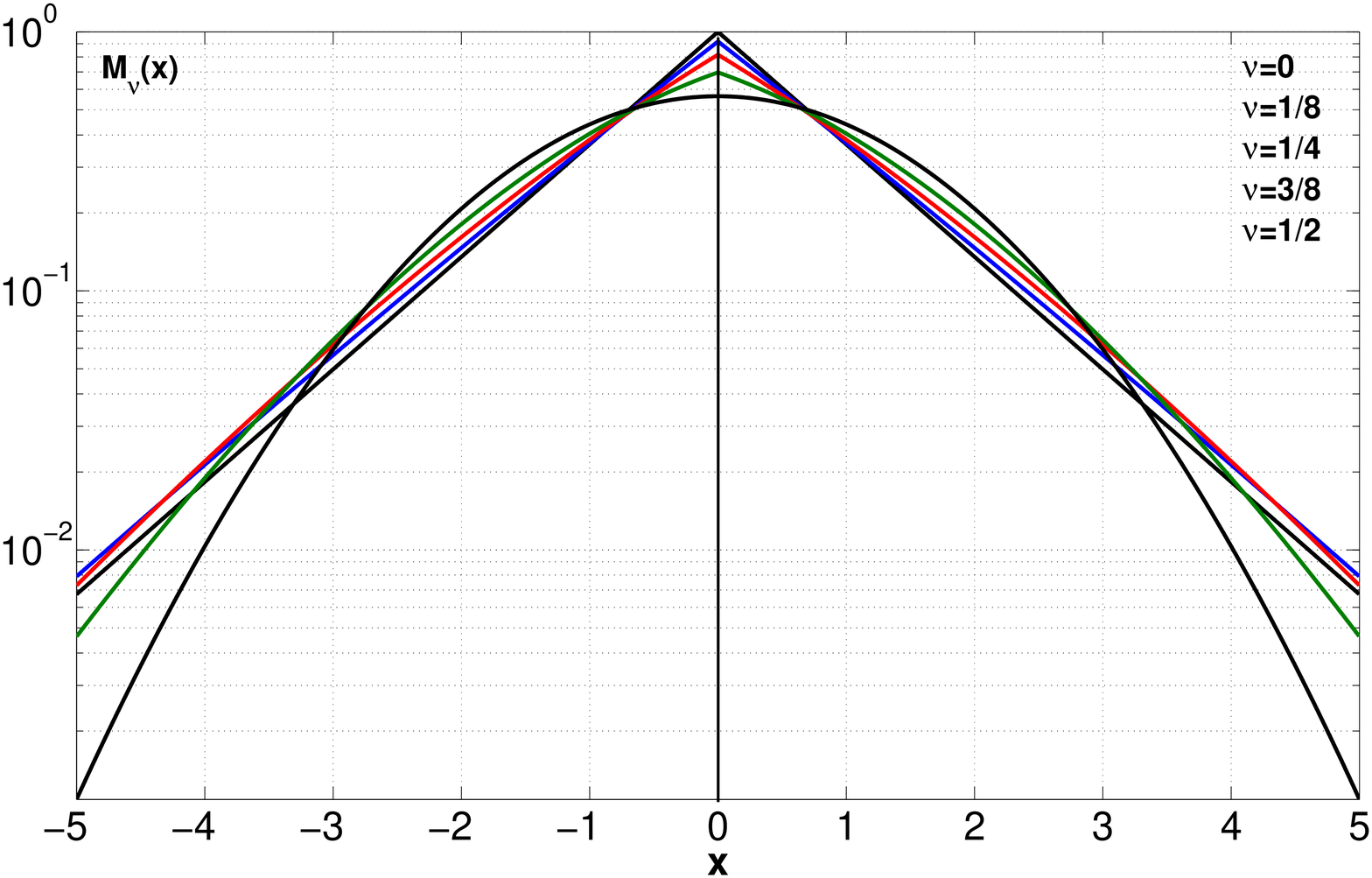}
\end{center}
 \vskip -0.6truecm
 \caption{Plots  of the  symmetric $M_\nu$-Wright function 
  with $\nu=0, 1/8, 1/4, 3/8, 1/2$ 
 for $-5\le x \le 5$;
 left: linear scale,
 right: logarithmic scale.
 \label{fig:F.1}}
\end{figure}
 \begin{figure}[ht!]
\begin{center}
\includegraphics[width=.45\textwidth]{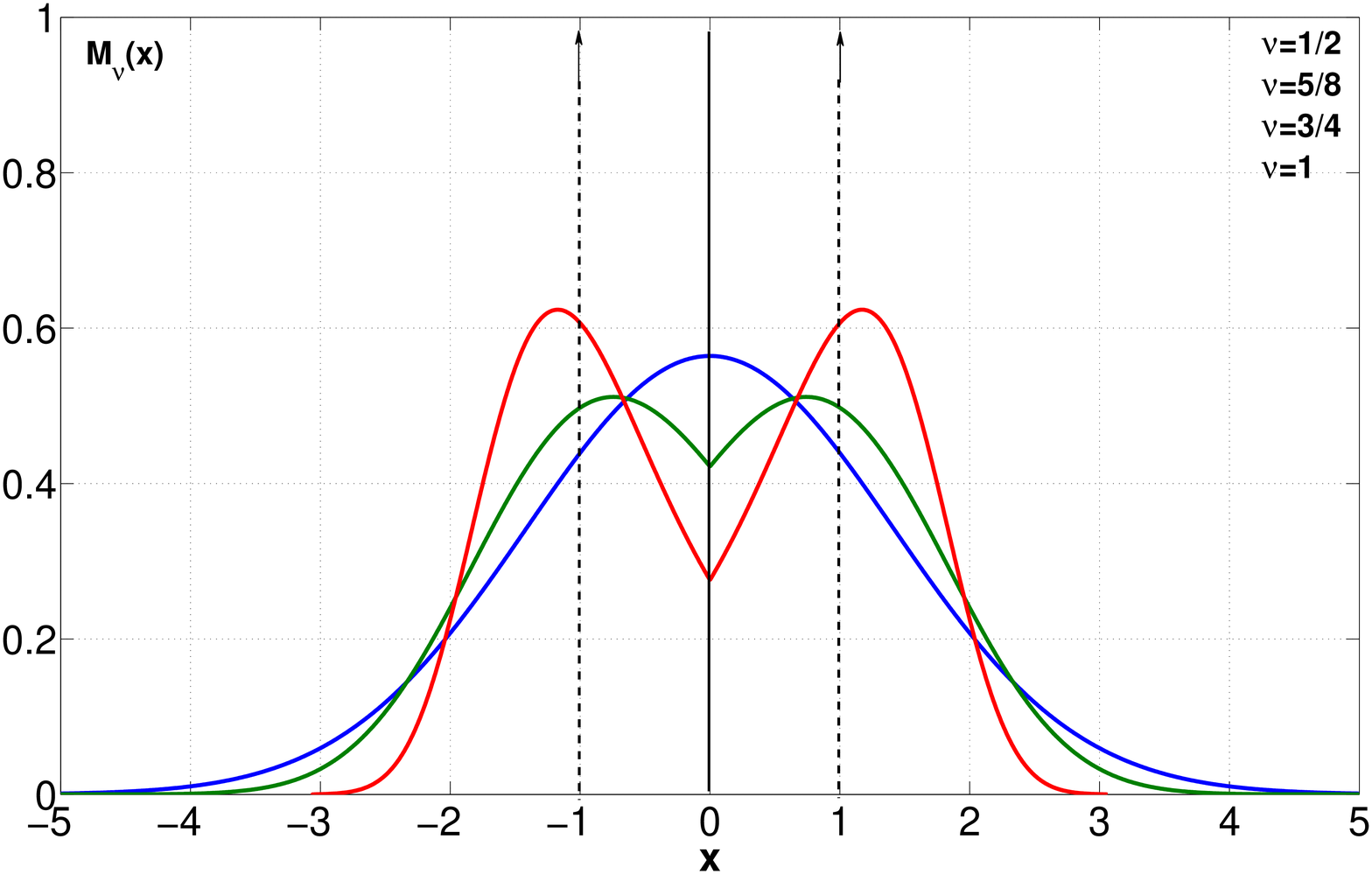}
\includegraphics[width=.45\textwidth]{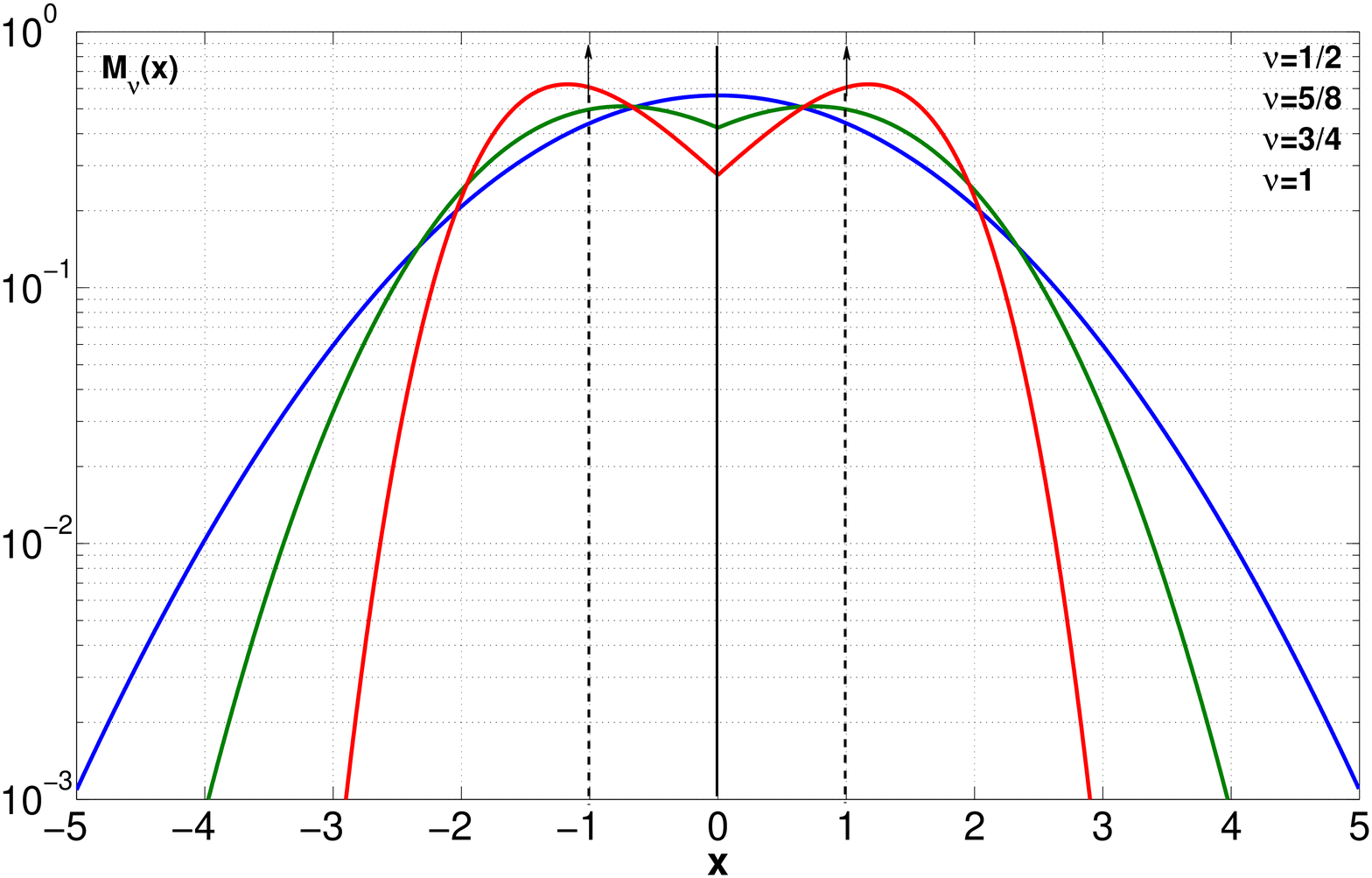}
\end{center}
 \vskip -0.6truecm
 \caption{Plots  of the  symmetric $M$-Wright function
  with $\nu=1/2\,,\, 5/8\,,\, 3/4\,, \,1$ 
 for $-5\le x \le 5$;
 left: linear scale;
 right: logarithmic scale.
 \label{fig:F.2}}
 \end{figure}
 \paragraph{The $\MM$-Wright function in two variables.}
 \vsp
In view of the time-fractional diffusion processes that will be considered in the next Sections, 
it is worthwhile to introduce the 
function  in two variables 
  $$\MM_\nu(x,t):= t^{-\nu}\, M_\nu(xt^{-\nu})\,,\q 0<\nu < 1\,,\q x,t \in \RR^+ \,,\eqno(4.14)$$
  which defines a  spatial probability density in $x$ evolving in 
  time $t$ with self-similarity exponent $H=\nu$.
  Of course for $x\in \RR$ we have to consider the symmetric version obtained from 
  (4.14) multiplying by $1/2$ and replacing $x$ by $|x|$.
  \vsp 
   Hereafter we provide
  a list of the main properties of this function, 
  which can be derived from Laplace and Fourier transforms of the corresponding $M$-Wright function 
  in one variable.  
 \vsp
   From Eq. (4.2) we   derive the Laplace transform of $\MM_\nu(x,t)$ with respect to $t \in\RR^+$,   
   $$\L\left\{\MM_\nu (x,t); t\to s \right\}= s^{\nu-1}\, \e^{\ds \, -xs^\nu}\,.\eqno(4.15)$$
    From Eq. (4.6) we   derive the Laplace transform of $\MM_\nu(x,t)$ with respect to $x\in \RR^+$,
	$$\L\left\{\MM_\nu(x,t); x\to s \right\}= E_{\nu}\left( -s t^\nu \right)\,.\eqno(4.16)$$
     From Eq. (4.10) we   derive the Fourier transform of $\MM_\nu(|x|,t)$ with respect to $x\in \RR$,
	$$\F\left\{\MM_\nu(|x|,t); x\to \kappa \right\}= 2E_{2\nu}\left( -\kappa^2 t^\nu \right)\,.\eqno(4.17)$$
Moreover,  using  the Mellin transform,  
  Mainardi et al. \cite{Mainardi-Pagnini-Gorenflo_FCAA03} derived 
the following integral formula,  
 $$\MM_\nu(x,t)= \int_0^\infty \MM_\lambda(x,\tau)\, \MM_\mu(\tau,t)\, d\tau\,,\q \nu = \lambda \mu\,.
 \eqno(4.18)$$
 Special cases of the  $\MM$-Wright function are simply derived for $\nu=1/2$ and $\nu=1/3$ from
 the corresponding ones in the complex domain, see Eqs. (3.10)-(3.11).
 We devote particular attention to the case $\nu=1/2$  for which we get  from (4.4)
 the Gaussian density in $\RR$,
 $$ \rec{2}\,\MM_{1/2}(|x|,t) = \rec{2\sqrt{\pi}t^{1/2}}\, \e^{\ds\, -x^2/(4t)} \,. \eqno(4.19)$$
 For the limiting case $\nu=1$ we obtain
 $$\rec{2}\, \MM_1 (|x|,t) = \rec{2} \left[\delta(x-t)+ \delta(x+t)\right]\,.\eqno(4.20)$$
\section{Fractional diffusion equations}
Let us now consider a variety of diffusion-like equations starting from the standard diffusion equation
whose fundamental solutions are expressed in terms of the $M$-Wright function 
depending on  space and time variables. 
The two variables, however, turn out to be related through a self-similarity condition.   
\paragraph{The standard diffusion equation.}
\vsp
The standard diffusion equation for the field $u(x,t)$ with initial condition
$u(x,0)=u_0(x)$ is 
$$
\frac{\partial u}{\partial t}= K_1 \, \frac{\partial^2 u}{\partial x^2} 
\quad -\infty < x < \infty \,, \;t \ge 0 \,,
\eqno(5.1)
\label{standard}
$$
where $K_1$ is a suitable diffusion coefficient of dimensions $[K_1]= [L]^2 [T]^{-1}= cm^2/sec$. 
This initial-boundary value problem can be easily shown to be   equivalent to the Volterra integral equation 
$$ u(x,t)= u_0(x) + K_1 \int_0^t \frac{\d^2 u(x,\tau)}{\dx^2}\,d\tau\,.\eqno(5.2)$$
  It is well known that the fundamental solution (usually refereed  as the {\it Green function}),
which is the solution corresponding  to $u_0(x)= \delta(x)$, is the Gaussian probability density 
evolving in time  with variance (mean square displacement) proportional to time. 
In our notation we hve: 
$$ \G_1(x,t) = \frac{1}{2 \sqrt{\pi K_1}\, t^{1/2}} \,\e^{\, \ds -x^2/(4K_1t)}\, , \eqno(5.3)$$
$$\sigma_1^2(t) := \int_{-\infty}^{+\infty}\!\! x^2 \, \G_1(x,t)\, dx =  2K_1t\,. \eqno(5.4)$$
This variance law characterizes the process of {\it normal diffusion}
as it emerges from  Einstein's  approach to   {\it Brownian motion} ($Bm$), 
see \eg \cite{Sokolov-Klafter_CHAOS05}.
\vsp
In view of future developments, we  rewrite the Green function in terms of the $M$-Wright function
by recalling Eq. (3.10), that is, 
$$   \G_1(x,t) = \rec{2} \rec{\sqrt{K_1}\, t^{1/2}}
   \,M_{1/2}\left( \frac{|x|}{\sqrt{K_1}\,t^{1/2}}\right)\, . \eqno(5.5)$$
From the self-similarity of the Green function in (5.3) or (5.5) 
we are led to write    
$$\G_1(x,t) =  \rec{\sqrt{K_1}\, t^{H}}\, \G_1 \left(\frac{|x|}{\sqrt{K_1}\,t^{H}},1 \right)\,, \eqno(5.6)$$
where $H=1/2$ is the  similarity (or Hurst) exponent and  $\xi = |x|/(\sqrt{K_1}\,t^{1/2})$ acts as 
the similarity variable.  We refer to the one-variable function $\G_1(\xi)$ as the reduced Green function.

\paragraph{The stretched-time standard diffusion equation.}
\vsp
Let us now stretch  the time variable in Eq. (5.1) by replacing $t$ with $t^\alpha$ where $0<\alpha < 2$.
We have
$$  \frac{\d u}{\d (t^\alpha)}\ = K_\alpha\, \frac{\d^2 u}{\dx^2}\,, 
 \q  -\infty<x<+\infty\,,\; t\ge 0\,, \eqno(5.7)$$
 where $K_\alpha$ is a sort of stretched diffusion coefficient of dimensions
 $[K_\alpha]= [L]^2 [T]^{-\alpha}= cm^2/sec^\alpha$. 
 It is easy to recognize that such equation is akin to  the standard diffusion equation 
 but with a diffusion coefficient depending on time,
 $K_1(t)= \alpha t^{\alpha-1}\, K_\alpha$.
  In fact, using the rule 
$$
\frac{\partial}{\partial t^\alpha}=\frac{1}{\alpha t^{\alpha-1}} \frac{\partial}{\partial t} \,,
$$
we have 
 $$  \frac{\d u}{\dt} =  \alpha t^{\alpha-1}\, K_\alpha\, \frac{\d^2 u}{\dx^2}\,, 
 \q  -\infty<x<+\infty\,,\; t\ge 0\,.
  \eqno(5.8)$$
\\
The integral form corresponding to Eqs. (5.7)-(5.8) reads
$$u(x,t)=u_0(x)+ \alpha K_\alpha \,
\int_0^t \frac{\partial^2 u(x,\tau)}{\partial x^2} \tau^{\alpha-1} \, d\tau \,.\eqno(5.9)$$
The corresponding fundamental solution is the stretched-time Gaussian
  $$ \G_\alpha(x,t) = \frac{1}{2 \sqrt{\pi K_\alpha}\, t^{\alpha/2}}\, \e^{\, \ds -x^2/(4K_\alpha t^\alpha)}
   = \rec{2} \rec{\sqrt {K_\alpha} \,t^{\alpha/2}}\,M_{1/2}\left( 
   \frac{|x|}{\sqrt{K_\alpha}\, t^{\alpha/2}}\right)\, . 
   \eqno(5.10) $$
 The corresponding variance 
 $$\sigma _\alpha ^2(t) := \int_{-\infty}^{+\infty}\!\! x^2 \, \G_\alpha(x,t)\, dx =  
 2K_\alpha t^\alpha\,, \eqno(5.11)$$
  is characteristic 
 of a general process of {\it anomalous  diffusion}, precisely of {\it slow diffusion} for $0<\alpha<1$,
 and  {\it fast diffusion} for $1<\alpha < 2$. 

\paragraph{The time-fractional diffusion equation.}
\vsp 
In  literature there exist two forms of the time-fractional diffusion equation of a single order, one
with  Riemann-Liouvile derivative and one with  Caputo derivative
These forms are equivalent if we refer to the standard initial condition $u(x,0)= u_0(x)$, 
as shown in \cite{Mainardi-Pagnini-Gorenflo_AMC07}.
\vsp
Taking a real number $\beta \in (0,1)$, the time-fractional diffusion equation 
of order $\beta$ in the Riemann-Liouville sense reads  
     $$
\frac{\partial u}{\partial t}= {K_\beta} \,  {D^{1-\beta}_t} \,\frac{\partial^2 u}{\partial x^2} \,,
\label{fractional} \eqno(5.12)
$$
whereas in the Caputo sense reads
$$_*D^{\beta}_t u = {K_\beta} \,  \frac{\partial^2 u}{\partial x^2} \,,
\eqno(5.13)$$
where $K_\beta$ is a sort of fractional diffusion coefficient of dimensions 
$[K_\beta]= [L]^2  [T]^{-\beta}= cm^2/sec^\beta$.
Like  for diffusion equations of integer order (5.1) and (5.7)-(5.8), we consider the equivalent 
integral equation  corresponding to our fractional diffusion equations (5.12)-(5.13), 
$$
u(x,t)=u_0(x)+K_\beta \, \frac{1}{\Gamma(\beta)} \int_0^t 
(t-\tau)^{\beta-1} \,\frac{\partial^2 u(x,\tau)}{\partial x^2} \, d\tau \,.
\eqno(5.14)$$
The Green function $\mathcal{G}_\beta(x,t)$ for  the  equivalent Eqs. (5.12)-(5.14) 
can be expressed, also in this case, in terms of the $M$-Wright function, as shown 
in Appendix by adopting two different approaches, as follows:  
$$
\mathcal{G}_\beta(x,t)= \frac{1}{2}
\rec{\sqrt{K_\beta}\, t^{\beta/2}}\, M_{\beta/2}
\left(\frac{|x|}{\sqrt{K_\beta}\, t^{\beta/2}}\right) \,.
\eqno(5.15) $$
The corresponding variance can be  promptly obtained from the general formula (5.5) for the absolute  moment
of the $M$-Wright function. In fact, using (5.5) and (5.15) and after an obvious change of variable,
we obtain  
 $$\sigma _\beta ^2(t) := \int_{-\infty}^{+\infty}\!\! x^2 \, \G_\beta(x,t)\, dx =  
   \frac{2}{\Gamma(\beta+1)}\,  K_\beta \,t^\beta\,. \eqno(5.16)$$
 As a consequence, for $0<\beta <1$ the variance is consistent 
 with a  process of {\it slow diffusion} with similarity exponent $H=\beta/2$.
 For further reading on  time-fractional diffusion equations and their solutions  the reader is referred,
 among others,  to
 \cite{Mainardi_CISM97,Mainardi_LUMAPA01,Mainardi-Pagnini_AMC03} and 
 \cite{Saichev_97}, \cite{SchneiderWyss_89}. 
\paragraph{The stretched time-fractional diffusion equation.}
\vsp 
In the fractional diffusion equation (5.12),
let us stretch  the time variable   by replacing
$t$ with $t^{\alpha/\beta}$ where $0<\alpha < 2$ and $0<\beta \le 1$.  
We have
$$
\frac{\partial u}{\partial t^{\alpha/\beta}}= K_{\alpha\,\beta}
\, D^{1-\beta}_{t^{\alpha/\beta}} \, \frac{\partial^2 u}{\partial x^2} \,,
\eqno(5.17)$$
namely
$$\frac{\partial u}{\partial t}=\frac{\alpha}{\beta} t^{\alpha/\beta-1} \,
K_{\alpha\,\beta} \, D^{1-\beta}_{t^{\alpha/\beta}}
\,\frac{\partial^2 u}{\partial x^2} \,,
\label{streched-fractional}
\eqno(5.18)$$
where $K_{\alpha\,\beta}$ is a sort of stretched diffusion coefficient of dimensions
 $[K_{\alpha\, \beta}]= [L]^2 [T]^{-\alpha}= cm^2/sec^\alpha$ that reduces to $K_\alpha$ if $\beta=1$
 and to $K_\beta$ if $\alpha= \beta$.
Integration  of  Eq. (5.18) gives 
the corresponding integral equation \cite{Mura-Pagnini_JPhysA08}  
$$
u(x,t)=u_0(x)+K_{\alpha\,\beta}\, \frac{1}{\Gamma(\beta)}\frac{\alpha}{\beta}\,
\int_0^t \tau^{\alpha/\beta-1}\, (t^{\alpha/\beta}-\tau^{\alpha/\beta})^{\beta-1}
\,\frac{\partial^2 u(x,\tau)}{\partial x^2} \, d\tau \,,
\eqno(5.19)$$
whose Green function $\mathcal{G}_{\alpha \,\beta}(x,t)$ is
$$
\mathcal{G}_{\alpha \,\beta}(x,t)= \rec{2}\,\rec{\sqrt{K_{\alpha\,\beta}}\, t^{\alpha/2}}\, 
M_{\beta/2}\left(\frac{|x|}{\sqrt{K_{\alpha \beta}}\, t^{\alpha/2}}\right) \,,
\eqno(5.20)
$$
with variance 
$$\sigma _{\alpha,\beta} ^2(t) := \int_{-\infty}^{+\infty}\!\! x^2 \, \G_{\alpha,\beta}(x,t)\, dx =  
   \frac{2}{\Gamma(\beta+1)}\,  K _{\alpha\,\beta} \,t^\alpha\,. \eqno(5.21)$$
 As a consequence, the resulting process turns  out to be self-similar
 with Hurst exponent $H=\alpha/2$ and  a variance law consistent both  with slow diffusion if  $0<\alpha <1$
 and fast diffusion if $1<\alpha < 2$.
 We note that the parameter $\beta$ does  explicitly enter in the  variance law (5.21) only  
  in the determination of the  multiplicative constant. 
 \vsp
 It is straightforward to note that the evolution equations of this process  
 reduce to those for time-fractional diffusion if $\alpha=\beta < 1$,
 for   stretched diffusion  if $\alpha \ne 1$ and $\beta =1$, 
 and finally to standard diffusion if $\alpha=\beta=1$. 
 
  
\section{Fractional diffusion processes with stationary increments}
  
We have seen that any Green function associated to the diffusion-like equations considered in the previous Section 
can be interpreted as the time-evolving one-point $pdf$ of certain self-similar stochastic processes. 
However, in general, it is not possible to define a {\it unique} (self-similar) stochastic process 
because the determination of any multi-point probability distribution is required, 
see e.g. \cite{Mura-Taqqu-Mainardi_PhysicaA08}.
 \\ \\
\noindent 
In other words, starting from a master equation which describes the dynamic evolution of 
a probability density function $f(x,t)$, it is always possible to define an equivalence class of 
stochastic processes with the same marginal density function $f(x,t)$. 
All these processes provide suitable stochastic representations for the starting equation. 
It is clear that additional requirements may be stated in order to 
uniquely select  
the probabilistic model.
\\ \\
For instance, considering Eq. (5.18), the additional requirement of stationary increments, 
as shown by Mura et al.,
see \cite{Mura_PhD08,Mura-Mainardi_ITSF09,Mura-Pagnini_JPhysA08,Mura-Taqqu-Mainardi_PhysicaA08}, 
can lead to a class $\{B_{\alpha,\beta}(t),\; t\ge 0 \}$, called {\it ``generalized'' grey Brownian motion} 
($ggBm$), 
which, {\it by construction}, is made up of self-similar processes with stationary increments 
and Hurst exponent $H=\alpha/2$. Thus  $\{B_{\alpha,\beta}(t),\; t\ge 0 \}$ 
is a  special class of $H\!-\!sssi$ processes\footnote{
According to a common terminology, $H\!-\!sssi$ stands 
 for $H$-self-similar-stationary-increments, see for details \cite{Taqqu_REV02}.},
which provide 
non-Markovian stochastic models  for anomalous diffusion,
both of slow type ($0<\alpha<1$) and fast type ($1<\alpha<2$). 
\\ \\
The $ggBm$ includes some well known processes, so that it defines an interesting general 
theoretical framework.
The fractional Brownian motion ($fBm$) appears for $\beta=1$ and is associated with Eq. (5.7); 
the grey Brownian motion ($gBm$), defined by Schneider \cite{Schneider_GN90a,Schneider_GN90b}, 
corresponds to the choice $\alpha=\beta$, with $0<\beta<1$, and is associated to  Eqs. (5.12), (5.13) or (5.14); 
finally, the standard Brownian motion ($Bm$) is recovered by setting $\alpha=\beta=1$ 
being  associated to Eq. (5.1).
We should note that only  in the particular case of $Bm$ the corresponding process is Markovian.    
\\ \\
In Figure 3 we present a diagram that allows to identify the elements of the $ggBm$ class.
The  top region $1<\alpha <2$ corresponds to the domain of fast diffusion with    
{\it long-range dependence}\footnote{A self-similar process with stationary increments  is said to 
possess long-range dependence  
if the  autocorrelation function of the increments tends to zero like a power function and such that 
it does not result integrable, see for details  \cite{Taqqu_REV02}.}. 
 In this domain the increments of the process $B_{\alpha,\beta}(t)$ are positively 
 correlated,  so that the trajectories tend to be more regular ({\it persistent}).
 It should be noted that long-range dependence is associated to a non-Markovian process
 which exhibits long-memory  properties. 
The horizontal line $\alpha=1$ corresponds to  processes  
with 
uncorrelated increments, which  model various phenomena of normal diffusion. 
For $\alpha =\beta=1$ we recover the Gaussian process of the standard Brownian motion.
 The Gaussian process of the fractional Brownian motion is identified by the vertical  line $\beta=1$.
The bottom region $0<\alpha<1$ corresponds to the domain of slow diffusion.
The increments of the corresponding process $B_{\alpha,\beta}(t)$  
turn out to be negatively correlated and this implies that the trajectories are 
strongly irregular   
({\it anti-persistent motion});  the increments form a stationary process which 
does not exhibit long-range dependence.
Finally, the  diagonal line ($\alpha=\beta$) represents the Schneider grey Brownian motion
($gBm$).
\begin{figure}[ht!]
\begin{center}
\includegraphics[width=0.9\textwidth]{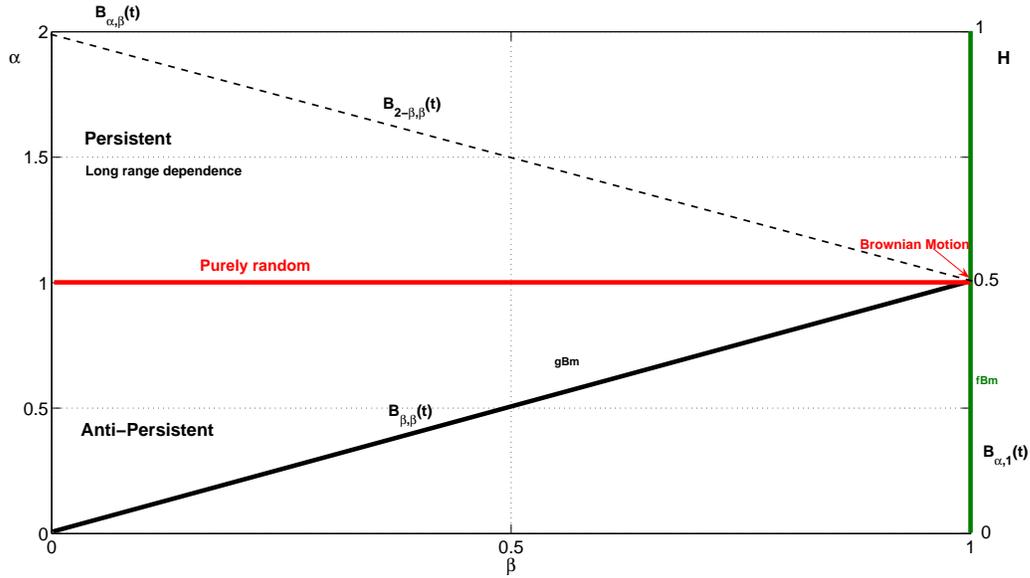}
 \end{center}
 \vskip -0.5truecm
 \caption{Parametric class of generalized grey Brownian motion 
 \label{fig:F.3}}
\end{figure}

\noindent
Here we want to define the $ggBm$ by making use of the Kolmogorov extension theorem 
and  the properties of the $M$-Wright function.
According to Mura and Pagnini \cite{Mura-Pagnini_JPhysA08},
the generalized grey Brownian motion $B_{\alpha,\beta}(t)$ 
is a stochastic process defined in a certain probability space 
such that its finite-dimensional distributions are given by
$$
f_{\alpha,\beta}(x_1,x_2,\dots,x_n;
\gamma_{\alpha,\beta})=\displaystyle\frac{(2\pi)^{-\frac{n-1}{2}}}{\sqrt{2\Gamma(1+\beta)^n
\det{\gamma_{\alpha,\beta}}}}\int_{0}^{\infty}\frac{1}{\tau^{n/2}}M_{1/2}\left(\frac{\xi}{\tau^{1/2}}\right)
M_\beta(\tau)d\tau,
\eqno(6.1)
$$
with
$$
\xi=\left(2\Gamma(1+\beta)^{-1}\sum_{i,j=1}^{n}x_i{\gamma_{\alpha,\beta}}^{-1}(t_i,t_j)x_j\right)^{1/2},
\eqno(6.2)$$
 and covariance matrix 
$$
\gamma_{\alpha,\beta}(t_i,t_j)=
\frac{1}{\Gamma(1+\beta)}(t_i^{\alpha}+t_j^{\alpha}-|t_i-t_j|^{\alpha}),\;\; i,j=1,\dots,n\,.
\eqno (6.3)$$
The covariance matrix (6.3) 
 characterizes the typical dependence structure of a self-similar process 
with stationary increments  and  Hurst exponent $H=\alpha/2$, see \eg \cite{Taqqu_REV02}.
\\ \\
Using Eq. (4.18), for $n=1$, Eq. (6.1) reduces to:
$$
f_{\alpha,\beta}(x,t)=
\displaystyle\frac{1}{\sqrt{4t^{\alpha}}}
\int_{0}^{\infty}\MM_{1/2}\left(|x|t^{-\alpha/2},
\tau\right)\MM_\beta(\tau,1)\,d\tau =\frac{1}{2}t^{-\alpha/2}M_{\beta/2}(|x|t^{-\alpha/2})\,.
\eqno(6.4) 
$$
This means that the  marginal density function of the $ggBm$ is indeed the fundamental solution (5.20) of 
Eqs. (5.17)-(5.18) with $K_{\alpha\beta}=1$.
Moreover, because
$M_1(\tau)=\delta(\tau-1)$,
for $\beta=1$, putting $\gamma _{\alpha,1}\equiv \gamma _\alpha$, we have that Eq.~(6.1) 
provides the Gaussian distribution of the fractional Brownian motion,
$$
f_{\alpha,1}(x_1,x_2,\dots,x_n;\gamma_{\alpha,1})=
\displaystyle\frac{(2\pi)^{-\frac{n-1}{2}}}{\sqrt{2\det{\gamma_{\alpha}}}}
M_{1/2}\left(\left(2\sum_{i,j=1}^{n}x_i\gamma_{\alpha}^{-1}(t_i,t_j)x_j\right)^{1/2}\right)\,,
\eqno(6.5)
$$
which finally reduces to the standard Gaussian distribution of Brownian motion as $\alpha=1$.
\\ \\
By the definition used above, it is clear that,  
fixed $\beta$, $B_{\alpha,\beta}(t)$ is  characterized  
  only by its   covariance structure, as shown by Mura et al.
  \cite{Mura-Mainardi_ITSF09}, \cite{Mura-Pagnini_JPhysA08}.  
 In other words, the $ggBm$, which is not Gaussian in general, is an example of a process defined only 
 through its first and second moments, which indeed is a remarkable property of Gaussian processes. 
 Consequently, the $ggBm$ appears to be  a direct generalization of Gaussian processes, 
 in the same way as the $M$-Wright function is  a generalization of the Gaussian function.

\section{Concluding discussion}

In this review  paper we have surveyed a quite general approach
to derive models for anomalous diffusion 
 based on a family of time-fractional
diffusion equations depending on two parameters $\alpha \in (0,2)$, $\beta \in (0,1]$.
\\ \\
The unifying topic of this analysis is the so-called $M$-Wright function
by which the fundamental solutions of these equations are expressed.
Such function is shown to exhibit fundamental analytical properties that
were properly used in recent papers for  characterizing and simulating a general class 
of self-similar stochastic processes with stationary increments including
fractional Brownian motion and grey Brownian motion.
\\ \\
In this respect, the $M$-Wright function emerges to be a natural generalization
of the Gaussian density to model diffusion processes, covering both 
slow and fast anomalous diffusion and including non-Markovian property.
In particular, it turns out to be the main function for the special 
 $H-sssi$ class of stochastic processes (which are self-similar with stationary increments)
governed by a master equation of fractional type.
 
 \section*{Acknowledgments}
 This work has been carried out in the framework of the research  project 
{\it Fractional Calculus Modelling} (URL: {\tt www.fracalmo.org}).
The authors are grateful to V. Kiryakova, R. Gorenflo and the anonymous referees for useful comments.

\section*{Appendix A: The fundamental solution of the time-fractional diffusion equation}


The  fundamental solution $\G_\beta(x,t)$ for the time-fractional diffusion equation can be obtained
by applying in sequence the Fourier and Laplace transforms to  any form  chosen
among Eqs. (5.12)-(5.14) with the initial condition $\G_\beta(x,0^+)= u_0(x) = \delta(x)$.
Let us devote our attention to the integral form (5.14)
using non-dimensional variables by setting $K_\beta =1$ 
and adopting the notation $J_t^\beta$ for the fractional integral.
Then,  our  Cauchy problem  reads
$$\G_\beta(x,t)= \delta(x) + J_t^\beta \,\frac{\partial^2 \G_\beta}{\partial x^2}(x,t) 
\,.\eqno(A.1)$$
In  the Fourier-Laplace domain, after applying   formula (2.18) for the
Laplace transform   of the fractional integral 
and observing $\widehat \delta (\kappa ) \equiv 1$, see e.g. \cite{Gelfand-Shilov_BOOK64},
we get
 $$   \widehat{\widetilde{G_\beta}}(\kappa ,s) = \rec{s}
     -\frac{\kappa^2}{s^\beta}\,
   \widehat{\widetilde{G_\beta}}(\kappa ,s) \,,
$$
from which  
$$  \widehat{\widetilde{\G_\beta}}(\kappa ,s)
   =  \frac{ s^{\beta -1}}{s^\beta + \kappa ^2 }\,, \q 0<\beta \le 1\,, \q
    \Re (s) > 0\,,\; \kappa \in \RR\,. \eqno(A.2)$$
To determine the  Green function $\G_\beta(x,t)$
in the space-time domain we can follow two
alternative  strategies related to the
order in carrying out the inversions in (A.2).
\\
(S1) : invert  the Fourier transform
getting $\widetilde{\G_\beta} (x,s)$
   and   then invert the remaining  Laplace transform;
\\
(S2) : invert  the Laplace transform getting $\widehat{G_\beta} (\kappa ,t)$
and then invert the remaining  Fourier transform.
\vsp\noindent
{\it Strategy (S1):} Recalling the Fourier transform pair
$$ \frac{a}{b +\kappa^2} \, \Fdiv \, \frac{a}{2 b^{1/2}}\, \e^{\ds\, -|x|b^{1/2}}\,, \q a,b>0\,,
\eqno(A.3)$$
and setting $a=s^{\beta-1}$, $b=s^\beta$, we get
$$\widetilde{\G_\beta} (x,s)= \rec{2}s^{\beta/2-1}\, \e^{\ds\, -|x|s^{\beta/2}}\,.\eqno(A.4)$$
 {\it Strategy (S2):} Recalling the Laplace transform pair
 $$ \frac{s^{\beta-1}}{s^\beta +c} \, \Ldiv \, E_\beta(-c t^\beta)\,, \q c>0\,,\eqno(A.5)$$
and setting $c=\kappa^2$, we  have
$$ \widehat{G_\beta} (\kappa ,t) = E_\beta(-\kappa^2 t^\beta)\,.\eqno(A.6)$$
Both strategies lead to the result
$$ \G_\beta(x,t) = \rec{2}\MM_{\beta/2}(|x|,t) = \rec{2}\, t^{-\beta/2}\, 
M_{\beta/2}\left(\frac{|x|}{t^{\beta/2}} \right)\,, \eqno(A.7)$$
consistent with Eq. (5.15). Here we have used  the $\MM$-Wright function, introduced in Section 4,
and its properties  related to the Laplace transform pair (4.15) for inverting  (A.4) and 
the Fourier transform pair (4.17) for inverting  (A.6).


\section*{Appendix B: The fundamental solution of the time-fractional drift equation}

Let us finally note that the $M$-Wright function does appear also in the fundamental solution of 
the time-fractional drift equation. 
Writing this equation in non-dimensional form 
and adopting  the Caputo derivative we have 
$$_*D^{\beta}_t u(x,t) = -  \frac{\partial }{\partial x} u(x,t) \,,\q -\infty<x< +\infty\,, \; t\ge 0\,, \eqno(B.1)
$$
where $0<\beta<1$ and $u(x,0^+) = u_0(x)$.
When $u_0(x)= \delta(x)$ we  obtain the fundamental solution (Green function) that we denote by
$\G_\beta^*(x,t)$.
Following the approach of Appendix A, we show that 
$$\G_\beta^*(x,t) = 
 \left\{
  \begin{array}{ll}
  {\ds t^{-\beta}\, M_\beta\left(\frac{x}{t^\beta}\right)}\,, & x>0\,,\\
  0\,, & x<0 \,,
  \end{array}
  \right.
  \eqno(B.2) 
$$
that for $\beta=1$ reduces to the right running pulse $\delta(x-t)$ for $x>0$.
\vsp
In  the Fourier-Laplace domain, after applying   formula (2.19) for the
Laplace transform   of the Caputo fractional derivative 
and observing $\widehat \delta (\kappa ) \equiv 1$, see e.g. \cite{Gelfand-Shilov_BOOK64},
we get
 $$   s^\beta\,\widehat{\widetilde{G_\beta^*}}(\kappa ,s)- s^{\beta-1} = +i\kappa\,
   \widehat{\widetilde{G_\beta^*}}(\kappa ,s) \,,
$$
from which  
$$  \widehat{\widetilde{\G_\beta^*}}(\kappa ,s)
   =  \frac{ s^{\beta -1}}{s^\beta -i\kappa  }\,, \q 0<\beta \le 1\,, \q
    \Re (s) > 0\,,\; \kappa \in \RR\,. \eqno(B.3)$$
Like in Appendix A, to determine the  Green function $\G_\beta^*(x,t)$
in the space-time domain we can follow two
alternative  strategies related to the
order in carrying out the inversions in (B.3).
\\
(S1) : invert  the Fourier transform
getting $\widetilde{\G_\beta} (x,s)$
   and   then invert the remaining  Laplace transform;
\\
(S2) : invert  the Laplace transform getting $\widehat{G_\beta^*} (\kappa ,t)$
and then invert the remaining  Fourier transform.
\vsp\noindent
{\it Strategy (S1):} Recalling the Fourier transform pair
$$ \frac{a}{b -i\kappa} \, \Fdiv \, \frac{a}{ b}\, \e^{\ds\, -xb}\,, \q a,b>0\,,\; x>0\,,
\eqno(B.4)$$
and setting $a=s^{\beta-1}$, $b=s^\beta$, we get
$$\widetilde{\G_\beta^*} (x,s)= s^{\beta-1}\, \e^{\ds\, -x s^{\beta}}\,.\eqno(B.5)$$
 {\it Strategy (S2):} Recalling the Laplace transform pair
 $$ \frac{s^{\beta-1}}{s^\beta +c} \, \Ldiv \, E_\beta(-c t^\beta)\,, \q c>0\,,\eqno(B.6)$$
and setting $c=-i\kappa$, we  have
$$ \widehat{G_\beta^*} (\kappa ,t) = E_\beta(i\kappa t^\beta)\,.\eqno(B.7)$$
Both strategies lead to the result (B.2).
\vsp
In view of Eq. (4.1) we also recall that the $M$-Wright function is related to the unilateral 
{\it extremal stable density} of index $\beta$. Then,
using our notation stated in  \cite{Mainardi_LUMAPA01} for stable densities,
 we write our Green function as 
$$\G_\beta^*(x,t) = \frac{t}{\beta} \,x^{-1-1/\beta}\, L_\beta^{-\beta}\left(t x^{-1/\beta}\right)\,, 
\eqno(B.8)$$
 \vsp
 To conclude this Appendix let us briefly discuss the above results in view of their relevance 
 in fractional diffusion processes following the recent paper by 
 Gorenflo and Mainardi \cite{GorMai_BAD-HONNEF08}.
Equation (B.1) describes the evolving sojourn probability density of the positively oriented time-fractional 
drift process of a particle, starting in the origin at the instant zero. It has been 
derived in \cite{GorMai_BAD-HONNEF08}  
as a properly scaled limit for the evolution of the counting number of the Mittag-Leffler 
renewal process (the fractional Poisson process). It can be given in several forms, and often it 
is cited as the {\it subordinator} (producing the operational time from the physical time) for 
{\it space-time-fractional  diffusion} as in the form  (B.8). For more details see
\cite{GorMaiViv_CSF07}, where simulations of space-time-fractional diffusion processes have been considered
as composed by time-fractional and space-fractional diffusion processes. 
\vsp
This analysis can be compared to that described with a different language in  papers by Meerschaert et al. 
 \cite{Meerschaert-et-al_PRE02,Meerschaert-Scheffler_JAP04}. 
Recently, a more exhaustive analysis has been given by Gorenflo \cite{Gorenflo_PALA09}.  


\end{document}